\documentclass[aps,tightenlines,twocolumn,groupaddress]{revtex4}
\usepackage{graphicx}% Include figure files
\usepackage{dcolumn}% Align table columns on decimal point
\usepackage{bm}% bold math
\usepackage{csquotes}
\usepackage{epigraph}
\usepackage{amssymb}
\usepackage{color,graphicx}
\usepackage{lmodern}
\usepackage{amsmath}
\usepackage{amsbsy}
\usepackage{amsthm}
\usepackage{float}
\usepackage{bbm}

\begin{document}

%\title{On the Description of Earth Dynamics Based on Bohmian Trajectories}
\title{Determination of classical behaviour of the Earth for large quantum numbers using quantum guiding equation}

\author{Ali Soltanmanesh}
\email[]{soltanmanesh@ch.sharif.edu}
\author{Afshin Shafiee}
\email[Corresponding Author:~]{shafiee@sharif.edu}
\affiliation{Research Group on Foundations of Quantum Theory and Information,
Department of Chemistry, Sharif University of Technology
P.O.Box 11365-9516, Tehran, Iran}
\affiliation{School of Physics, Institute for Research in Fundamental Sciences (IPM), P.O.Box 19395-5531, Tehran, Iran}

\begin{abstract}
For quantum systems, we expect to see classical behavior at the limit of large quantum numbers. Hence, we apply the Bohmian approach to describe the Earth evolution around the Sun. We obtain possible trajectories of the Earth system with different initial conditions which converge to a certain stable orbit after a given time, known as the Kepler orbit. The trajectories are resulted from the guiding equation $p=\nabla S$ in Bohmian mechanics which relates the momentum of the system to the phase part of the wave function. Except at some special situations, Bohmian trajectories are not Newtonian in character. We show that the classic behavior of the Earth can be interpreted as the consequence of the guiding equation at the limit of large quantum numbers. 
\end{abstract}

%%insert keywords separated by comma using \keywords{words}

%%include \pacs{number} to print the PACS number

\maketitle

%%%%%%%%%%%%%%%%%%%%%%%%%%%%%%%%%

\section{Introduction}

Quantum mechanics works exceedingly well in all practical applications. No examples of conflict between its predictions and experiments are known. The main problem arises when we emphasize on macroscopic systems which behave classically. The transition of quantum to classical mechanics is an unsolved fundamental problem for years. Most physicists believe that macroscopic systems are quantum mechanical in nature. Therefore, the way they behave classically is not clear yet. In the most general sense, the correspondence principle determines the guideline of scientific theories. However, not only the latest achievements require to make predictions that earlier studies were inadequate to address, but also they need to confirm the correct predictions of the previous theories \cite{Kee}. In this issue, Bohmian mechanics which comes up with trajectories for quantum systems can be an appropriate option for the development of other studies describing the micro-world.

Bohmian mechanics is a deterministic and distinctly non-Newtonian reformulation of quantum mechanics which the wave function itself is responsible for guiding the motion of the particles \cite{Durr}.  However, despite indiscernibility in results with the standard quantum mechanics, it differs in explanations \cite{Hol}. Through self-experiences and the classical description of nature, we comprehend macro-systems with trajectories. Therefore, there is no way to understand the quantum-classical transition, until we reach an explanation for classical trajectories based on the quantum description. Since Bohmian mechanics keeps the quantum mechanics results based on its casual implementation, it can potentially provide a proper connection between the \linebreak quantum-classical domain. In this regard, many efforts have been done to extend the Bohmian description in a variety of physical problems such as deriving relativistic quantum potential \cite{Rah}, following the Bohmian approach in non-linear Klein-Gordon equation \cite{Gol}, The extension of a classical motion of a constrained particle to the quantum domain considering Bohm's view of quantum mechanics \cite{You}, and space-time transformation for the propagator in Bohm theory \cite{Til}. It should be noted that the trajectories which Bohmian mechanics implies are entirely different from the Newtonian trajectories so that the theory has its particular description \cite{Durr2}. Interestingly, in this regard, periodic orbits have been, since Kepler, considered as the key concept for describing and understanding classical dynamics which, since Bohr, could be gainful in apprehending the quantum-classical transition \cite{Flo}.

Describing the Sun-Earth dynamics was always major problem in physics \cite{Gut}.  The first illustration of Earth's orbit was computed by Lagrange (1781,1782), and improved by Pontecoulant (1834), Agassiz (1840) and others. Since then, many works have been done to modify the theory and ameliorate the results \cite{Las}. In recent times, many quantum research studies have been done to shed light on the problem. As an instance, Fl\"othmann and Welge showed the classical dynamics of electron motion of Hydrogen atom under the cross magnetic-electric field, in connection with the gravitational field of the Sun-Earth problem \cite{Flo}. Also, Battista and Esposito studied the effect of quantum corrections to the Newtonian potential for the evaluation of equilibrium points \cite{Bat}.

The real issue still has been remained unsolved. However, the dynamics of the famous classical Earth-Sun problem has unanswered questions. Meanwhile, the quantum aspects of the problem have been considered unsolved in general \cite{Gutz, Holl, Bet, Nie}. Nevertheless, both quantum and classical features of the model appear in all subfields of physics \cite{Car, Tan, Jen, Jon, Ran, Rob, Che, Yah, Cui, Cas}. Regarding classical Sun-Earth dynamics, many efforts have been made to describe the system in a quantum fashion. David Keeports considered Earth as a quantum object and discussed its quantum properties compared to classical ones \cite{Kee}. Also, studies on quantum potential correction terms recently reported \cite{Ham,Nao,Har}.

In this paper, we are going to present the quantum aspects which is responsible for the Earth to behave classically.  In section 2, we present non-relativistic Hydrogen-like Hamiltonian for the Sun-Earth system to show that the Earth classical energy is dependent upon the principal and the angular momentum quantum numbers $n$ and $l$, respectively. In section 3, we discuss the role of the Bohmian mechanics and its guiding equation in the appearance of the Earth dynamics. Then, in the next section, we show as one of our main results, that how the high values of the magnetic quantum number $m$ affect the possible trajectories of the Earth via the Bohmian guiding equation. Later, we discussed the wave function. Finally, in the last section 6, we addressed the results of our work.

%%%%%%%%%%%%%%%%%%%%%%%%%%%%%%%%%

\section{Definition of Earth Hamiltonian}

We know the Earth as a classical object, nontheless there is a belief that the behavior of macroscopic systems, even Earth in the solar gravitational field, is based on their quantum mechanical nature. Therefore, different forms of Hamiltonians for the description of the Earth dynamics have been proposed in literature \cite{Kee,Gut,Nie}.  Here, we introduce a Hamiltonian which contains
two terms: (1) the hydrogen-like Hamiltonian H0 containing the kinetic and gravitational potential energy terms and (2) an additional energy term denoted by $K$:
\begin{align}
\label{eq1}
&H=H_0+K \\
\label{eq2}
&H_0=-\frac{\hbar^2}{2m_e}\nabla^2-\frac{Gm_sm_e}{r},
\end{align}
where, $G$ is the universal gravitational constant, $m_s$ and $m_e$ are the Sun and the Earth masses respectively.

 We included the additional energy term $K$ regarding the derived Hamiltonian from two-body problem post-Newtonian (PN) approximation. Accordingly,  $r^{-1}$, $r^{-2}$ and the lower order energy terms of $r$ are included in the expressions of 1PN, 2PN, and 3PN contributions to the Hamiltonian \cite{Dam}. Also, since the dynamics here based on the Hydrogen atom two-body problem, relativistic perturbations result in additional $r^0$, $r^{-1}$ and $r^{-2}$ correction terms to the kinetic energy \cite{Tow}. Moreover, an ensemble of trajectories which are solutions to the equations of motions corresponds to a well-known theorem of mechanics \cite{Boh2}. Thus, concerning the recent discussion, we defined the additional $r^0$, $r^{-1}$ and $r^{-2}$ energy terms, in which the yielded equations of motion are consistent with the classical description of the problem. Evidently, including the exact terms of 1PN, 2PN, and 3PN, correction terms in the Hamiltonian lead to a more accurate and also a more complicated solution. It should be considered as a real part of the Earth Hamiltonian $H$, without which a proper description of the Earth’s dynamics is not possible, even in a classical-like (Bohmian) approach.  We introduce the additional term $K$ in the upcoming section.

Following the Bohmian approach in the description of the problem \cite{Boh}, we consider $\psi(x,t)=Re^{-iS/\hbar}$ as an eigenfunction of the Hamiltonian. Thus, by dividing the real and the imaginary parts of the Schr\"odinger equation, one reaches the following equations
\begin{align}
\label{eq4}
&\frac{\partial S}{\partial t}+\frac{(\nabla S)^2}{2m_e}-\frac{\hbar^2}{2m_e}\frac{\nabla^2 R}{R}-\frac{Gm_sm_e}{r}+K=0 \\
\label{eq5}
&\frac{\partial R^2}{\partial t}+\nabla\cdot(\frac{R^2\nabla S}{m_e})=0.
\end{align}
The relation (\ref{eq4}) is the well-known quantum Hamilton-Jacobi equation. Keeports pointed out that how the Hydrogen-like energy of the Earth is related to its classical one \cite{Kee}. He showed if the Earth-Sun Hamiltonian is assumed similar to the Hydrogen atom, regardless of the additional term $K$, the energy levels could be calculated as $E_n=-G^2m_e^3m_s^2/2\hbar^2n^2$.  On the other hand, by considering the continuous limit of the Earth classical energy, he obtained
\begin{align}
\label{Energy}
 E_n&=-\frac{G^2m_e^3m_s^2}{2\hbar^2n^2}=-\frac{Gm_e^3m_s^2}{2\hbar^2l(l+1)}, 
 \end{align}
where $n$ and $l$ are the principal and the azimuthal quantum numbers. As is clear, he suggested that we should have $n=l\rightarrow \infty$. Also, Keeports showed that the Earth wave function is the same as the Hydrogen atom when the quantum numbers $n$ and $l$ tend to large values.

The Hydrogen-like Hamiltonian $H_0$, explains the classical energy of Earth very well, as we mentioned above. So, it is reasonable to assume that the energy value remains unchanged in this new framework. This point is crucial in analyzing the role of different terms in the Hamilton-Jacobi equation (\ref{eq4}). Moreover, the term $-\frac{\hbar^2}{2m_e}\frac{\nabla^2 R}{R}$ is negiligible due to the large mass of the Earth. Considering the Energy relation \eqref{Energy}, for the Hamiltonian \eqref{eq1} the term $(\nabla S)^2/2m_e+K$ should be equivalent with $(\nabla S)^2/2m_e$ for Hydrogen-like Hamiltonian ($H_0$). Accordingly, we can write the following expression in spherical coordinates
\begin{equation}
\label{eq7}
\frac{(\nabla S)^2}{2m_e}+K=\frac{m^2\hbar^2}{2m_er^2\sin^2\theta},
\end{equation}
where $m$ denotes the so-called magnetic quantum number.

The difference here from the system with \linebreak Hydrogen-like Hamiltonian is that both $r$ and $\theta$ are supposed to be time-dependent. Thereupon our first assumption here is that the energy of the Earth wave function is in a form similar to the same system with Hydrogen-like Hamiltonian. (For Hydrogen-like Earth-Sun system see \cite{Kee}).  The first assumption is an appropriate statement due to the fact that our explanation should be in a complete agreement with the classical results. Hence, we seek for an answer for the Schr\"odinger equation which achieves the energy \eqref{Energy}. Furthermore, we discuss the trajectories resulted from the wave function.
 
Hence, the study of the Earth in the Bohmian approach necessitates the accordance with quantum mechanical results, reaching the energy relation \eqref{Energy} in the same way. For the Hydrogen atom, the gradient of the phase function is $(\nabla S)^2/2m_e=\linebreak m^2\hbar^2/2m_er^2\sin^2\theta$ \cite{Hol}. So, the equation (\ref{eq4}) should have the same form as in the Hydrogen atom. This condition, however, leads to a new phase function and also provides a different dynamics afterward.

%%%%%%%%%%%%%%%%%%%%%%%%%%%%%%%%%%%%%

\section{Bohmian Dynamics of The Earth}

\begin{figure}[!t]
\centering{
\includegraphics[scale=0.41]{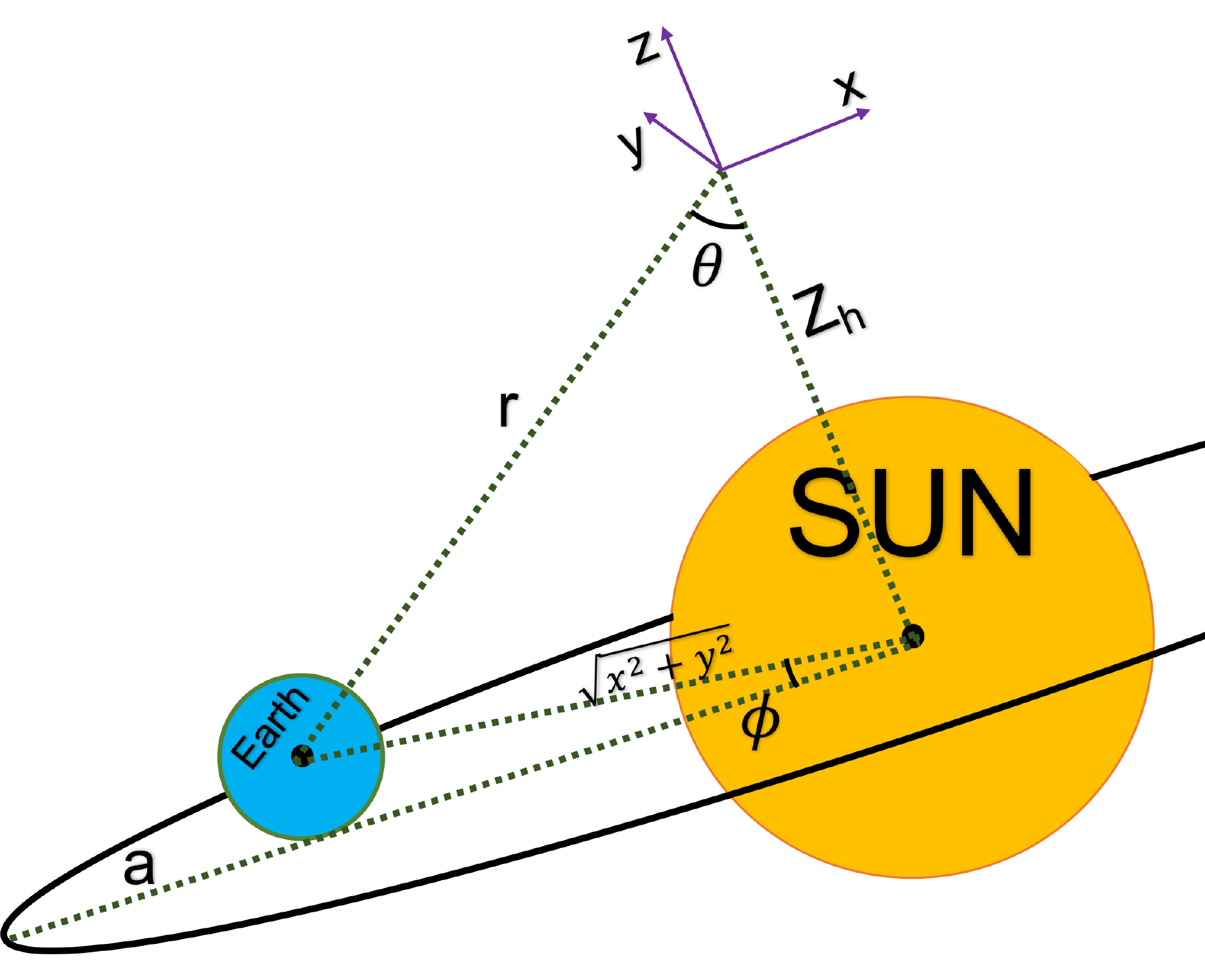}
}
\caption{Coordination axes, variables, and parameters in the Earth-Sun system.} \label{diagram}
\end{figure}

Keeping the energy unchanged as discussed in the previous section and according to the equation (\ref{eq7}), $K$ plays an essential role in the determination of the Earth trajectories. For the reasons formerly discussed, we define
\begin{equation}
\label{eqK}
K=\frac{Am_e}{2(r^2-Z_h^2)}-\mu m_e(\frac{1}{r}-\frac{1}{2a}),
\end{equation}
  where $a$ is the semi-major axis of Earth (see Table \ref{table1} for its value) and $\mu$ is defined as $\mu=G(m_s+m_e)\simeq Gm_s$, $A$ is a constant that we define later and $Z_h$ represents the $z$-axis of the coordinate system as shown in figure \ref{diagram}. Then in (\ref{eq7}) we have
\begin{align}
 \label{eq8}
 \frac{(\nabla S)^2}{2m_e}=&\frac{m^2\hbar^2}{2m_er^2\sin^2\theta}-\frac{Am_e}{2(r^2-Z_h^2)} \nonumber \\
 &+\frac{m_e\mu}{r}-\frac{m_e\mu}{2a}.
 \end{align}
Let us consider the $xy$-plane parallel to the plane of Earth's orbit. The $z$-axis is then perpendicular to this plane. So, the $z$-coordinate of the Earth position must be constant, \textit{i.e.},
 \begin{equation}
 \label{eq9}
 r\cos\theta=Z_h,
 \end{equation}
 where one can write
 \begin{equation}
 \label{eq10}
r^2\sin^2\theta=r^2-Z_h^2.
 \end{equation}
 According to the negligible changes in radius, $r$, and polar angle, $\theta$, during the Earth evolution, $Z_h$  almost considered as a constant as is obvious in figure \ref{diagram}. The value of $Z_h$ is pretty arbitrary for its value does not affect the dynamics. However, theoretically, it can be considered zero (the centre of the coordination system could be chosen as the centre of the Sun). Nevertheless, the equations are still not easy to solve. A reasonable choice for $Z_h$ is a constant in the same order of magnitude of Earth-Sun distance. In this case (according to figure \ref{diagram}) the radius, $r$ would be also in the same order of magnitude ($r=\sqrt{2}(x^2+y^2)^{1/2}$), thus the Hamiltonian \eqref{eq1} does not change due to the coordinate transformation. We would discuss this choice later through investigating the trajectories. As a suitable choice, we also define the constant $A$ in a way that the first two terms in R.H.S of (\ref{eq8}) cancel out each other. So, we adopt
 \begin{equation}
 \label{eq11}
 A=\frac{m^2\hbar^2}{m_e^2}.
 \end{equation}
 Consequently, in (\ref{eq8}) one can show that the Bohmian guiding equation is equal to
 \begin{equation}
 \label{eq12}
 (\nabla S)^2=m_e^2(\frac{2\mu}{r}-\frac{\mu}{a}).
 \end{equation}
Since the guiding equation $\nabla S=p$, where $p$ is the linear momentum, we obtain
 \begin{equation}
 \label{eq13}
 v^2=\frac{2\mu}{r}-\frac{\mu}{a}
 \end{equation}
 which is the well-known \textit{vis viva} relation in Newtonian mechanics. Regarding the equation (\ref{eq13}), velocity only depends on the distance between the Sun and the Earth.
 
In Bohmian mechanics, the dynamics of the wave function defined by the Schr\"odinger equation and the dynamics of the system determined by the guiding equation. There is a wrong general belief that the Bohmian mechanics become Newtonian for the macroscopic systems or the systems with large masses. This misleading concept happens when we ignore the guiding equation $\nabla S=p$. D\"urr and Teufel show that for many systems the Bohmian trajectories obtained from the guiding equation are different from the Newtonian ones \cite{Durr2}. The nature of Bohmian trajectories depends on the circumstances that the guiding equation is at work. We will show that, for the Sun-Earth system, the guiding equation leads to the trajectories which are classical in large values of quantum numbers.

 We must emphasize that as long as the wave function satisfies Schr\"odinger equation, that \linebreak $v=\nabla S/m$ and we have the ensemble with $\vert\psi\vert^2$ probability density, the Bohmian Results are quite identical with the one that quantum mechanics propose \cite{Boh2}, as it is in this manuscript. Indeed we use the results attained from the standard quantum dynamics \cite{Kee}, to extend it in Bohmian mechanics since the latter provides trajectories and leads to a better understanding of the problem in the macroscopic regime.

To show that our results here are entirely compatible with the standard quantum mechanics we need to discuss the three main aspects of the problem: Energy, wave function and the probability density.  As we argued before the dynamics here are based on the energy relation \eqref{Energy}, calculated by \linebreak Keeports, that in high values of $n$ are completely compatible with classical results \cite{Kee}. Moreover,  we previously computed the phase of the wave function $S$ using the quantum Hamilton-Jacobi equation \eqref{eq4}, regarding the energy relation \eqref{Energy} and the fact that the term  $-\frac{\hbar^2}{2m_e}\frac{\nabla^2 R}{R}$ (quantum potential) is negligible because of the mass of the Earth. Further in section 5, we will calculate the amplitude of the wavefunction $R$. Therefore, $\psi=Re^{-iS/\hbar}$ is the wave function of the Hamiltonian \eqref{eq1} with the energy eigenvalues equal to \eqref{Energy} regardless of the Copenhagen or Bohmian interpretation. Furthermore, we discuss in section 5 that the spatial probability density of Earth $\vert\psi\vert^2$ is entirely compatible with our calculated trajectories.  In the following, we are going to study the Bohmian solution of the problem and calculate the system trajectories.
%%%%%%%%%%%%%%%%%%%%%%%%%%%%%%%%%%%%%

\section{Earth Trajectories}

The electron in the Hydrogen atom has a circular trajectory in Bohmian mechanics with a constant radius in the $xy$-plane. Azimuthal angle $\phi(t)$ is the only parameter that changes with time, which has linear time-dependency \cite[pp. 148-153]{Hol}. The similarity of the Hydrogen system with the Earth one is a good reason to assume a similar relation for $\phi(t)$ in the Earth-Sun case. So, our second assumption here, is to consider the time dependency of the azimuthal angle $\phi(t)$ as same as the solution of the Hydrogen atom in Bohmian mechanics, albeit in its form. However, in this case, the radius $r$ and the polar angle $\theta$ are supposed to be time-dependent, contrary to the case of the Hydrogen atom. So we assume that
\begin{equation}
\label{eq14}
\phi(t)=\frac{m\hbar t}{m_er^2(t)\sin^2\theta(t)} +\phi_0.
\end{equation}
 This is an appropriate assumption according to the classical dynamics of the Earth's motion around the Sun. Accordingly, the time dependence of the azimuthal angle is in the order of $t$. The relation \eqref{eq14}, which is inspired from the Hydrogen atom dynamics, signifies the role of the magnetic quantum number $m$. Assuredly, the suggested relation needs to be in accordance with the Schr\"odinger equation. Furthermore, the important consequence of this choice is the direct effect of the quantum number $m$ on the dynamics of the system. The aforementioned is the main accomplishment of the Bohmian approach which appears in this problem. Our second assumption is not a surprise but is completely understandable. It is clear that the idea comes from the circular movement of the Earth around the Sun, and the equation (\ref{eq14}) is the simplest way to describe this movement regarding the theory.

\begin{table}[!t]
\centering
\caption{Important quantities and quantum numbers for Earth in quantum dynamics.  \label{table1}}

\begin{tabular}{c p{19ex} c}
\hline
Symbol & Quantity & Value \\
\hline
$r_{eq}$ & {\small The equilibrium distance between the Earth and the center of the coordinates} & $2.116\times10^{11}m$ \\[1ex]
$a$ & {\small The Earth semi-major axis} & $1.496\times10^{11}m$ \\[1ex]
$n$ & {\small Principal quantum number} & $2.524\times10^{74}$ \\[1ex]
$l$ & {\small Azimuthal quantum number} & $2.524\times10^{74}$ \\[1ex]
$m$  & {\small Magnetic quantum number} & $10^{73}$ \\[1ex]
$A$ & {\small $m^2\hbar^2/m_e^2$} & $10^{31}m^4s^{-2}$ \\[1ex]
$b$ & {\small $\hbar^2/Gm_e^2m_s$} & $2.348\times10^{-138}m$ \\[1ex]
\hline
\end{tabular}
\end{table}

Now, for the angular velocity $v_\phi$, we have
\begin{align}
\label{eq15}
v_{\phi}=r\sin\theta\dot{\phi}=\frac{m\hbar}{m_er^2\sin^2\theta}&-\frac{2m\hbar t}{m_er^3\sin^2\theta}\dot{r} \nonumber \\
&-\frac{2m\hbar t\cos\theta}{m_er^2sin^3\theta}\dot{\theta},
\end{align}
where $v^2=v_r^2+v_{\theta}^2+v_{\phi}^2$. Also using the equation (\ref{eq9}), one gets
\begin{equation}
\label{eq15a} 
v_{\theta}=r\dot{\theta}=\frac{Z_h\dot{r}}{r\sin\theta}.
\end{equation}
Using (\ref{eq15}), (\ref{eq15a}) and (\ref{eq13}), the time-dependent equation for $r$ is obtained as
\begin{align}
\label{eq16}
&(1+\frac{4At^2}{r^4\sin^2\theta}+\frac{Z_h^2}{r^2\sin^2\theta}+\frac{4AZ_h^4t^2}{r^8\sin^6\theta}+\frac{8AZ_h^2t^2}{r^6\sin^4\theta})\dot{r}^2 \nonumber \\
&-(\frac{4At}{r^3\sin^2\theta}+\frac{4AZ_h^2t}{r^5\sin^4\theta})\dot{r}+\frac{A}{r^2\sin^2\theta}-\frac{2\mu}{r}+\frac{\mu}{a}=0. \nonumber \\
\end{align}
Here, we need a solution for the radial velocity $\dot{r}$ to show the trajectories. To do this, we should know the magnitude of each term to make some approximations.

 The order of magnitude of the constant $A$ depends on the Earth mass, Planck constant, and the magnetic quantum number $m$. Therefore, its order must be determined owing to the Earth dynamics. As we know the Earth revolves around the Sun in a year. So, for the azimuthal angle $\phi(t)$, we have $\phi(\tau+1 \text{\footnotesize year})=\phi(\tau)+2\pi$. The radial time-dependency of the Earth motion is negligible, and its value remains nearly unchanged. Then, according to the definition of $\phi(t)$ in (\ref{eq14}), we can estimate the magnitude of the magnetic quantum number $m\approx10^{73}$. As we expect, the Earth dynamics appears in the large values of quantum numbers $l$ and $m$.
 
 Thus, from (\ref{eq11}) we have $A\approx10^{31}m^4s^{-2}$ and due to the equation (\ref{eq9}), the constant $Z_h$ should be in the order of the Earth-Sun distance. Now, let us assume that
 \begin{equation}
 \label{eq19}
 \dot{r}=\xi\sin\theta\sqrt{\frac{2\mu}{r}-\frac{\mu}{a}},
\end{equation}
 where $\xi$ is a dimensionless constant which will be determined later. Then, for the equation (\ref{eq16}), we have
 \begin{align}
 \label{eq18}
&\overbrace{(1+\frac{4AZ_h^2t^2}{r^6\sin^4\theta}(1+\sin^2\theta)+\frac{4At^2}{r^4})}^{\alpha}(\frac{2\mu}{r}-\frac{\mu}{a})\xi^2 \nonumber \\
 &-\frac{A\xi}{r^2\sin^2\theta}\underbrace{\frac{4t}{r\sin\theta}\sqrt{\frac{2\mu}{r}-\frac{\mu}{a}}}_{\beta}+\frac{A}{r^2\sin^2\theta} \nonumber \\
&-(\frac{2\mu}{r}-\frac{\mu}{a})=0.
 \end{align}
With respect to the Earth-Sun distance which is approximately constant and the fact that we are seeking the trajectories for the time domain of $10^7-10^8s$ magnitude, the coefficients $\alpha$ and $\beta$ are almost constant (about $10^1$) during the Earth revolution. Moreover, the fact that $\xi$ is assumed to be constant is due to the limited time domain we supposed here. For long time variations, $\xi$ is time-dependent. But then, the equation (\ref{eq18}) cannot be solved rigorously.

Regarding the equation (\ref{eq19}), the radial velocity $\dot{r}$ changes with time due to the time-dependency of $r(t)$ and $\theta(t)$. There is no analytical solution for this kind of nonlinear differential equation. Yet, by applying some physical and mathematical assumptions, the equation (\ref{eq19}) can give us proper trajectories, which are appropriate at the one year time domain. The method used here is to attain the following solution
\begin{align}
\label{S1}
r(t)&=\xi\frac{\sqrt{F^2(t)-\sqrt{F^2(t)(F^2(t)-4Z_h^2)}}}{\sqrt{2}} \\
\label{S2}
F(t)&=\frac{-B_2+C\tan[\frac{1}{2}C(t+\tau)]}{B_3}.
\end{align}
In (\ref{S2}) we define
\begin{align}
\label{S3}
&B_1:=B-\frac{\mu^2}{8B^3r_{eq}^2}+\frac{\mu}{2Br_{eq}}        &    &B_3:=\frac{\mu^2}{8B^3r_{eq}^4} \nonumber \\
&B_2:=\frac{\mu^2}{4B^3r_{eq}^3}-\frac{\mu}{2Br_{eq}^2}        &    &C:=\sqrt{-B_2^2+4B_1B_3} \nonumber \\
&B:=\sqrt{\frac{2\mu}{r_{eq}}-\frac{\mu}{a}}
\end{align}
where $\tau$ is a constant with time dimension. The details of calculating (\ref{S1}) from (\ref{eq19}) are given in Appendix. 

\begin{figure}[!t]
\centering{
\includegraphics[scale=0.45]{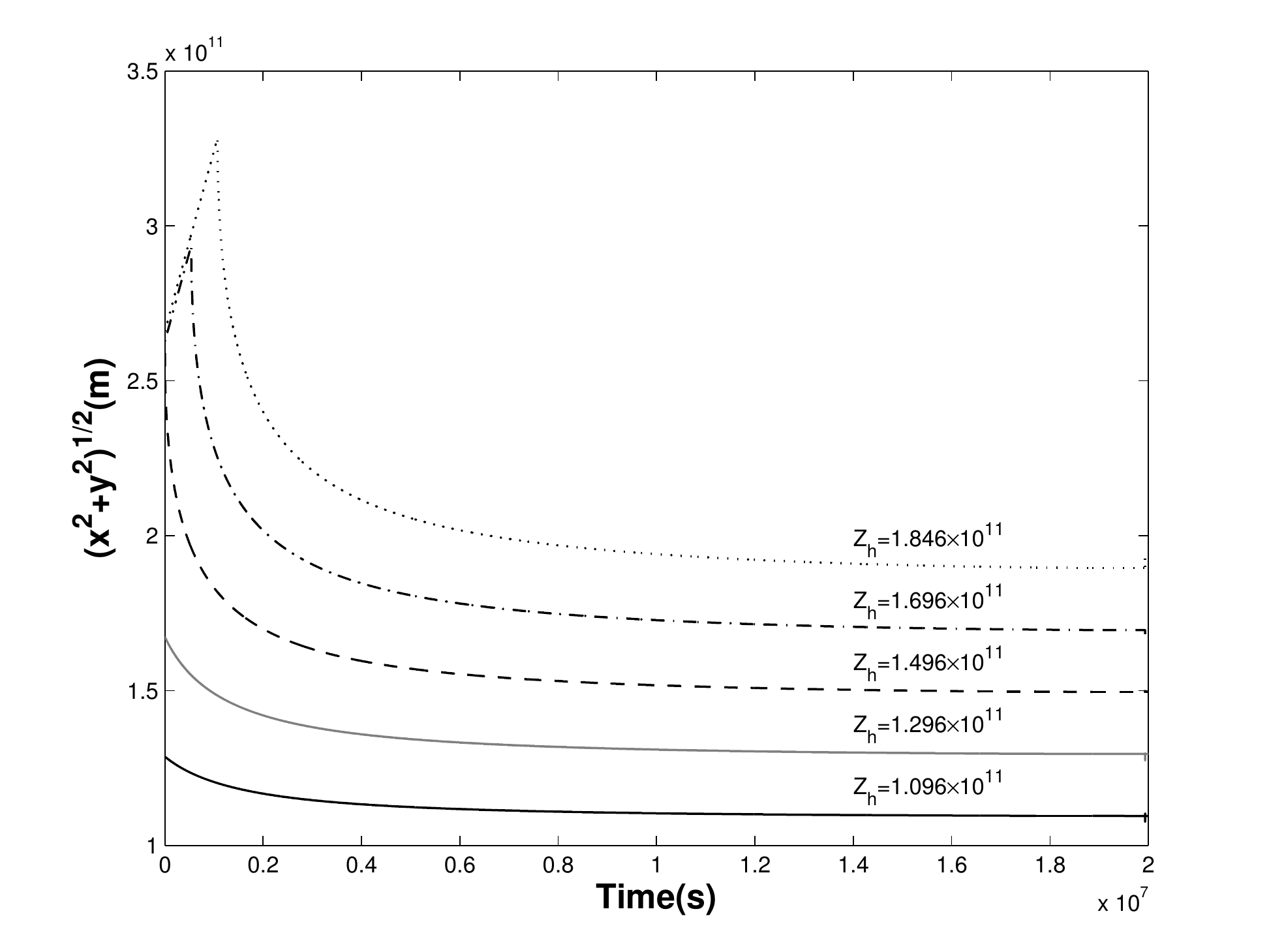}
}
\caption{The Earth-Sun distance versus the time for different values of the constant $Z_h$ and the typical value of $\xi=1.414$.} \label{fig1}
\end{figure}

According to the equations (\ref{eq9}), (\ref{eq14}) and (\ref{S1}) we know how $\phi(t)$, $\theta(t)$ and $r(t)$ vary with time. So it is now possible to draw the trajectories. However, some points should be made clear first. If we take a closer look at (\ref{S2}), we will find a periodic \textit{tangent} function with singularity points which means that our trajectories become discrete in these points. So, there is no continuous path prediction for all definite times $0\leq t<\infty$ in this model. We can only follow and draw the trajectories in limited time domains, e.g., for one year ($10^7-10^8s$). In each time interval (which can be extended even to several years), we can obtain closed cycles, demonstrating nearly the Earth orbit around the Sun.

The equation (\ref{S1}) shows us that $r$ decreases while time goes forward, then it reaches to an equilibrium value ($r_{eq}$). Such behavior sketched in figure \ref{fig1}.  The center of coordinates here is not located on Sun, but is displaced by $Z_h$. So, the radius $r$ is equal to $\sqrt{x^2+y^2+Z_h^2}$. Nevertheless, when we speak about the Earth-Sun distance, we mean $\sqrt{x^2+y^2}$, and as it discussed earlier $Z_h$, $r$ and Earth-Sun distance are in the same order of magnitude. As is represented in figure \ref{diagram}, $Z_h$ is just a displacement in center of coordinates, and the choice of its value is arbitrary due to the method of solution. According to figure \ref{fig1}, irrespective of the initial conditions, the Earth–Sun distance decreases rapidly and tends to an equilibrium value. In other words, different trajectories generated from different initial conditions converge to a particular equilibrium situation when time passes sufficiently.

\begin{figure}[!t]
\centering{
\includegraphics[scale=0.46]{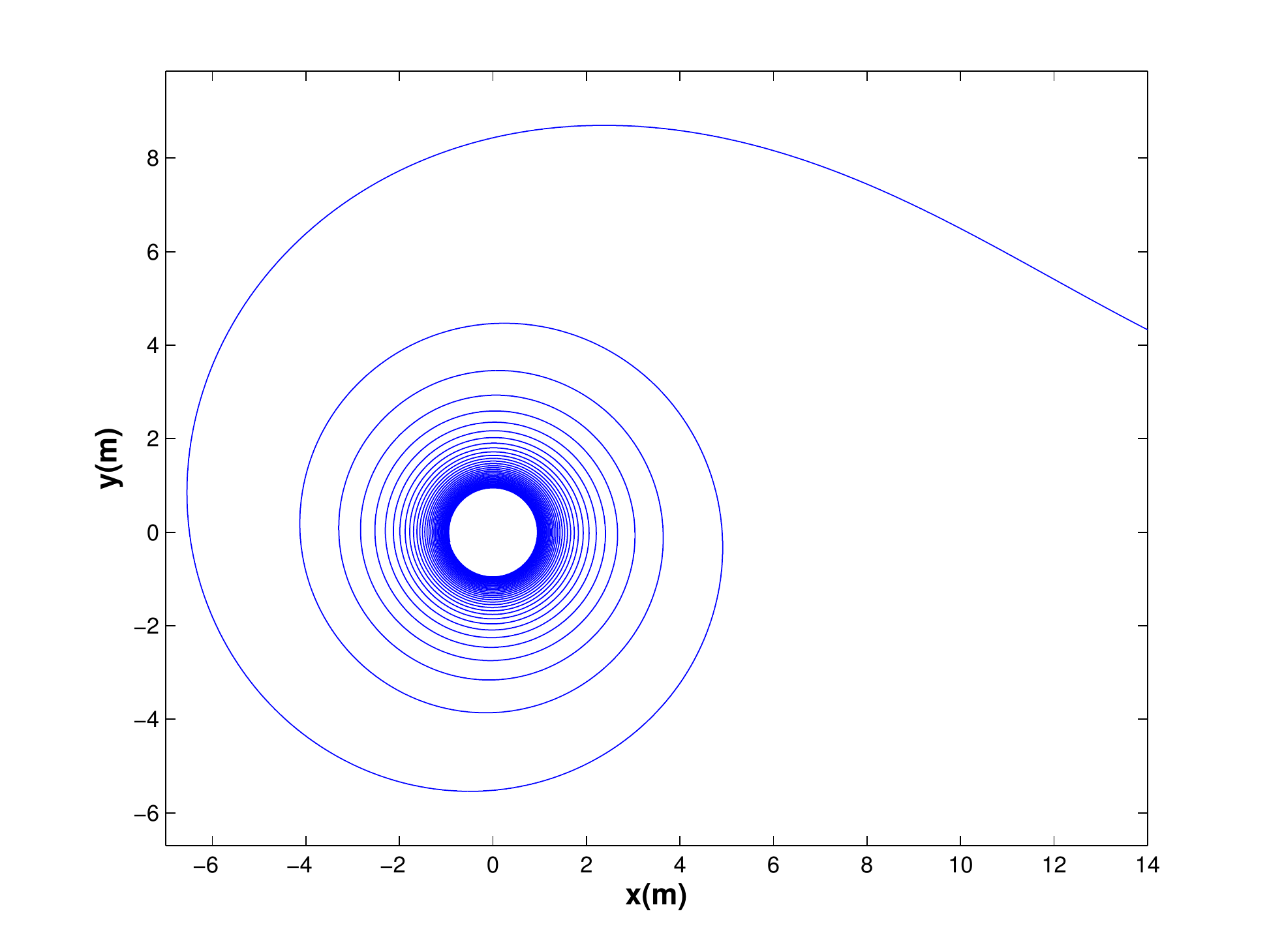}
}
\caption{The Earth trajectory around the Sun. After sufficient time (at the order of $10^7s$), the Earth-Sun distance tends to its classical equilibrium value of $1.496\times 10^{11}m$. The axes are $10^{11}$ times smaller.} \label{fig2}
\end{figure}

\begin{figure}[!t]
\centering{
\includegraphics[scale=0.8]{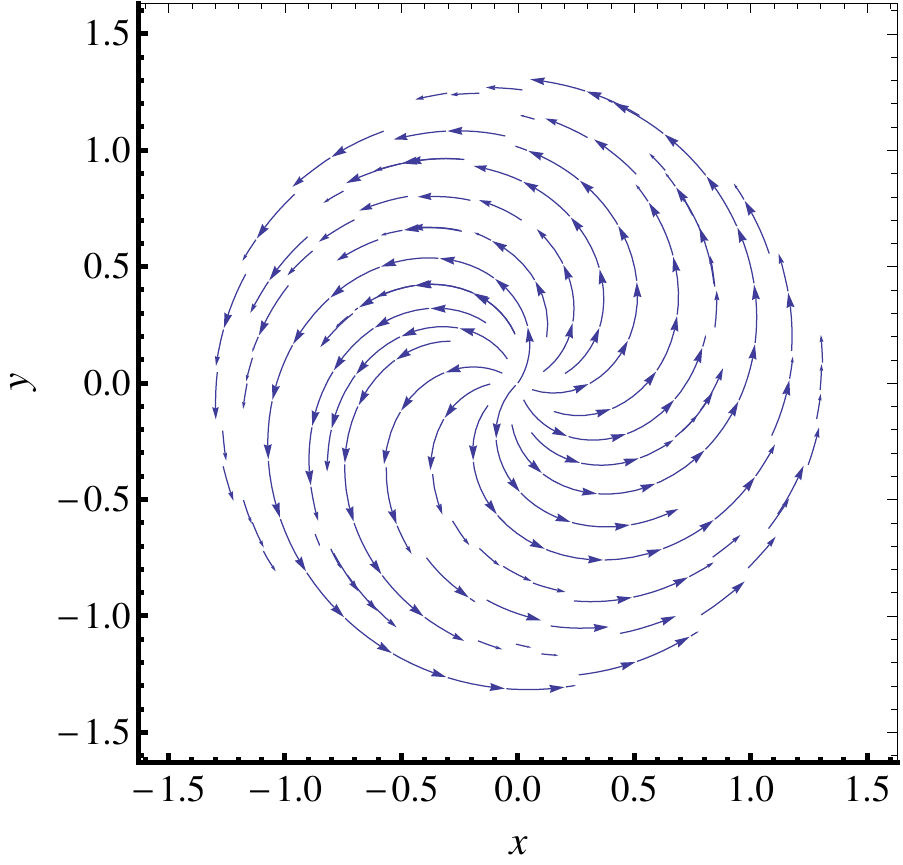}
}
\caption{The stream of Earth's vector field $(\dot{x},\dot{y})$ as a function of $x$ and $y$ coordinates in its final form.} \label{fig3}
\end{figure}

As we mentioned before, the trajectory equation (\ref{S1}) is not the exact answer of the differential equation (\ref{eq18}), though by an accurate choice for the values of $Z_h$ and $\xi$ which could be made by the classical data, the equation (\ref{S1}) would be practically an appropriate solution for the Earth trajectory in the Bohmian framework. For values of $Z_h=1.496\times10^{11}m$, $\phi_0=0$ and $\tau=0$, as the main initial conditions, figure \ref{fig2} shows how the Earth trajectory finally approaches to an stable closed cycle around the sun with an equilibrium distance $1.496\times10^{11}m$ ($r_{eq}=\sqrt{2}\times1.496\times10^{11}$). As an instance, we can assume that $\phi(t)=2\pi t$. Then, for the $x$ and $y$ components of the velocity we have 
\begin{align}
\label{eq20}
\dot{x}=&\frac{\sqrt{x^2+y^2}}{x^2+y^2+Z_h^2} \nonumber \\
&\times(1+\frac{Z_h^2}{x^2+y^2}x)\sqrt{\frac{2\mu}{\sqrt{x^2+y^2+Z_h^2}}-\frac{\mu}{a}}-2\pi y \\
\label{eq21}
\dot{y}=&\frac{\sqrt{x^2+y^2}}{x^2+y^2+Z_h^2} \nonumber \\
&\times(1+\frac{Z_h^2}{x^2+y^2}y)\sqrt{\frac{2\mu}{\sqrt{x^2+y^2+Z_h^2}}-\frac{\mu}{a}}+2\pi x.
\end{align}
Figure \ref{fig3} shows the stream of vector field $(\dot{x},\dot{y})$ which is described by the equations (\ref{eq20}) and (\ref{eq21}). As figure \ref{fig3} represents, the final equilibrium dynamics of the system is independent of initial conditions.

%%%%%%%%%%%%%%%%%%%%%%%%%%%%%%%%%

\section{The Wave Function}

\begin{figure}[!t]
\centering{
\includegraphics[scale=0.38]{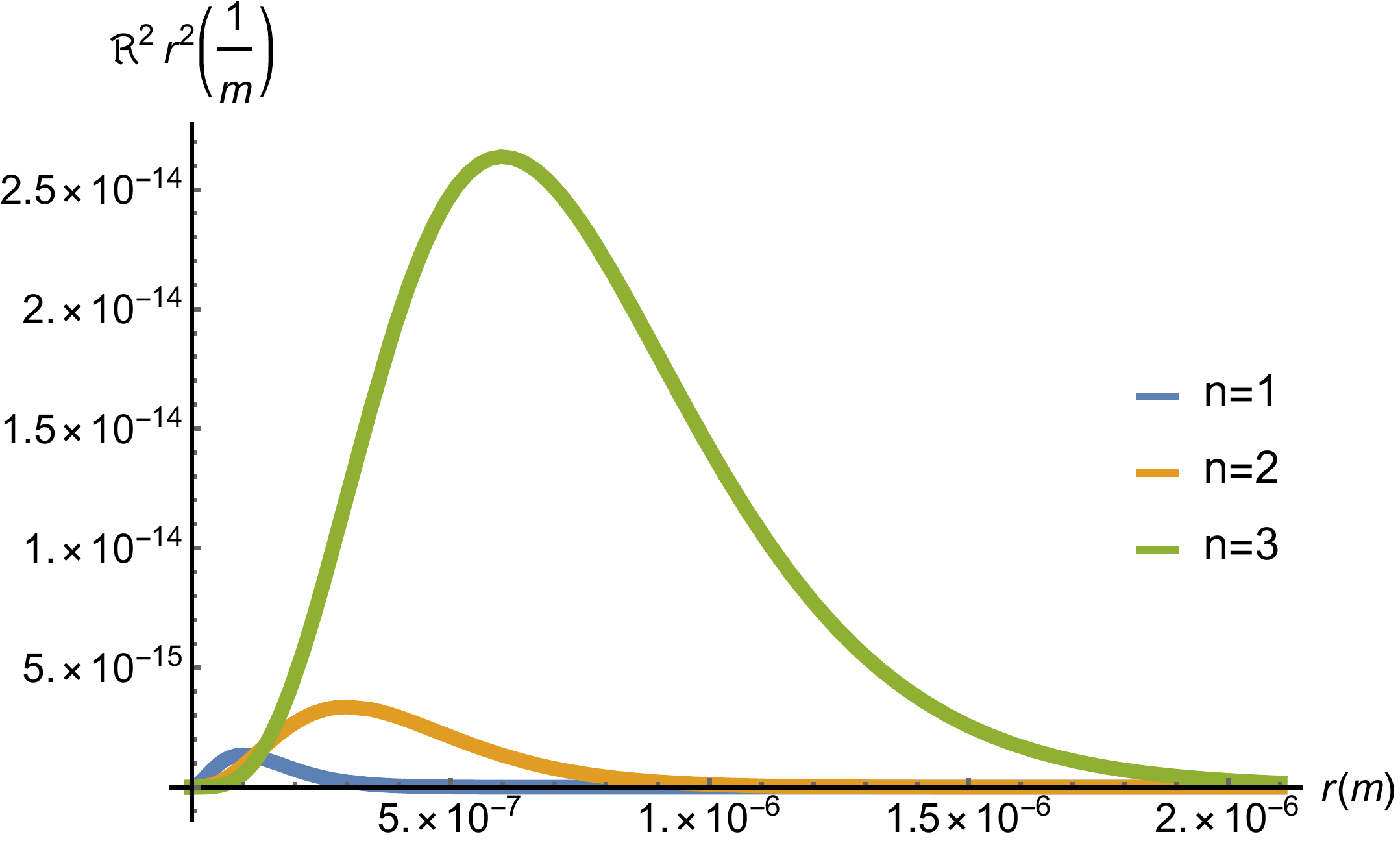}
}
\caption{Radial probability distribution $r^2\Re^2$ versus radius $r$ for $n=1,2,3$ for a typical value of $b=1\times 10^{-7}m$. For higher principal quantum number values, the probability density is more sharp around the most probable range. For values of $n$ and $b$ mentioned in TABLE 1 the Radial probability distribution tends to the Dirac delta function; i.e. $\delta(r-r_{mp})$. } \label{pd}
\end{figure}

So far we discussed the Earth dynamics and its trajectory. Here, we are going to investigate the Earth wave function. Let us note the time independent amplitude of the wave function, $R(r,\theta)$. Considering the equation (\ref{eq5}) we have
\begin{equation}
\label{eq22}
R^2\nabla\cdot\nabla S+\nabla S\cdot\nabla R^2=0.
\end{equation}
In spherical coordinates, one can write $\nabla S$ as
\begin{align}
\label{eq23}
\nabla S=&m_e\dot{r}\vec{\rho}_r+m_e\cot\theta\dot{r}\vec{\rho}_{\theta} \nonumber \\
&+\frac{\sqrt{A}}{r\sin\theta}(1-\frac{2\dot{r}t}{r}-\frac{2\dot{r}\cos^2\theta t}{r\sin^3\theta})\vec{\rho}_{\phi},
\end{align}
where $\dot{r}$ is defined in (\ref{eq19}). If we consider $R$ as $R(r,\theta)=\Re(r)\Theta(\theta)$, similar to the Hydrogen-like wave function, with the large values of $l$ and $m$ quantum numbers, one can write $\Theta(\theta)\propto\sin^m\theta$. According to $\Theta(\theta)$ relation, the function $\Theta^2$ as a probability distribution is approximately either $1$ for $\theta=\pi/2$ or zero for the other values of $\theta$ due to the large value of $m$. On the one hand, for $\Theta^2=0$ the relation (\ref{eq22}) is simply valid, but on the other hand, for $\Theta^2=1$ we have $R^2(r,\theta)\simeq\Re^2(r)$ and the validity of (\ref{eq22}) should be studied. By substituting $\nabla S$ from (\ref{eq23}) in (\ref{eq22}) and after some algebric calculations, one can show that
\begin{equation}
\label{eq24}
\frac{\partial \Re^2}{\partial r}=(\frac{a}{2ar-r^2}-\frac{1}{r})\Re^2.
\end{equation}
The equation (\ref{eq24}) imposes constraints on the amplitude of the wave function $R(r,\theta)$. Inspired by the Hydrogen atom wave function we assume that $\Re^2(r)$ is in the form of
\begin{align}
\label{amp}
\Re^2(r)=f(r)e^{-2r/bn},
\end{align}
in which $b=\hbar^2/Gm_e^2m_s$ and $f(r)=\sum_i^{n-1}C_ir^i$. Accordingly, regarding the large value of $n$ and the fact that $a$ is in the same order of magnitude as $r$, by substituting \eqref{amp} in \eqref{eq24} one can conclude that
\begin{align}
\label{eq25}
R^2(r,\theta)&=\Re^2(r)\Theta^2(\theta) \nonumber \\
&=C\frac{2^{n+2}}{n^{n+2}\Gamma(n+2)b^{n-1}}r^{n-1}e^{-2r/bn}\sin^m\theta 
\end{align}
where $C=b^{-3}$ is the normalization factor and for positive integers $\Gamma(m)=(m-1)!$. The radial probability density $r^2\Re(r)^2$ is a Gaussian-like distribution. We see that for higher values of the principal quantum number $n$, the function is more sharp around the most probable area with an exponential growth as is shown in figure \ref{pd}. By Maximizing the probability density one gets $r_{mp}=n(n+1)b/2$ as the most probable radius. Applying $b$ and equivallating the most probable radius with $r_{eq}$ from Table \ref{table1}, we obtain $n=2.524\times10^{74}$. Since the quantum number $n$ is quite large, the normalized probability distribution $r^2\Re^2(r)$ is perfectly sharp in $r_{mp}$. Accordingly, we can write  $r^2\Re^2(r)=\delta(r-r_{mp})$. Simply the provided trajectories are in a complete agreement with the probability density as is discussed formerly.  It has to be noted that the dynamics are settled regarding the conservation of the total probabilities, if all the trajectories exist for all times.

Here the wave function is an energy eigenstate as discussed before. However, with the large values of $n$, the differences between the energy levels are negligible, and the energy becomes continuous. Therefore, although the wave function is an energy eigenstate, the energy values are continual, in agreement with classical results.

%%%%%%%%%%%%%%%%%%%%%%%%%%%%%%%%%

\section{Conclusion}

Here, we have considered the Earth as a quantum object to show that the quantum guiding equation makes it behaves classically in the limit of large quantum numbers. Via the quantum formalism, the Earth dynamics determine by the Schr\"odinger equation. By describing the Sun-Earth system as a model of the Hydrogen atom, we introduced a new Hamiltonian with an additional kinetic energy term $K$.  To obtain the predicted quantum trajectory of the Earth, we used the guiding equation $p=\nabla S$ (see (\ref{eq12})) in the Bohmian regime.

Astonishingly, we see that the large quantum numbers directly ascertain the Earth dynamics. As Keeports has already shown \cite{Kee}, at large quantum numbers $n$ and $l$, the classic energy of the Earth obtained. Also, we showed here that the Earth dynamics depends on the high values of the quantum number $m$ which affects the Earth trajectory via the guiding equation (\ref{eq12}). The investigations led us to makes the well-known Newton's \textit{vis viva} equation (\ref{eq13}) as the Earth velocity, which leads to acceptable Bohmian trajectories, after some reasonable approximations.

Interestingly, the main result is independent of the initial conditions. The rotation distance of the Earth-Sun system decreases rapidly and tends to an equilibrium one. All trajectories with different initial conditions approach to a stable closed cycle, illustrating how Earth orbits around the Sun with negligible deviation. As a macrosystem, this is a magnificent achievement to approximately obtain the Earth trajectory in the Bohmian framework, which enables us to see the quantum footprints in a classical domain.

%%%%%%%%%%%%%%%%%%%%%%%%%%%%%%%%%

\section*{Acknowledgement}
The Authors would like to thank M. Koorepaz Mahmoodabadi for his assistance and useful comments on the nonlinear equations with closed cycles which improved our ideas on the subject.

%%%%%%%%%%%%%%%%%%%%%%%%%%%%%%%%%

\appendix
\section{}
The differential equation (\ref{eq19}) can be solved in any time domain in which $\xi$ is considered as a constant. One can write this equation as
\begin{equation}
\label{A1}
\frac{dr}{dt}=\xi\sin\theta\sqrt{\frac{2\mu}{r(t)}-\frac{\mu}{a}},
\end{equation}
where $\theta$ and $r$ are both time-dependent, but have no dependency on each other. For $\theta$ we have from (\ref{eq9})
\begin{equation}
\label{A2}
\sin\theta=\sqrt{1-\frac{Z_h^2}{r^2(t)}}.
\end{equation}
During the Earth rotation around the Sun, the time variations of $r$ are insignificant. With regard to (\ref{A2}) and noticing that the constant $Z_h$ has the same order of magnitude as $r$, one can neglect the time-dependency of $\theta$ too. Therefore, to solve the equation (\ref{A1}), we can nearly consider the term $\xi\sin\theta$ as a constant.

Defining the variable $q$ as
\begin{equation}
\label{A3}
q=r-r_{eq},
\end{equation}
where $r_{eq}$ represents the equilibrium distance, the equation (\ref{A1}) yields
\begin{equation}
\label{A4}
\frac{dq}{dt}=\xi\sin\theta\sqrt{\frac{2\mu}{r_{eq}(\frac{q}{r_{eq}}+1)}-\frac{\mu}{a}}.
\end{equation}
Hence, the variable $q$ represents the deviation of $r$ from the equilibrium distance $r_{eq}\simeq a$. The term $q/r_{eq}$ is small, so that one can expand $(1+q/r_{eq})^{-1}$ to obtain:
\begin{equation}
\label{A5}
\frac{dq}{dt}=\xi\sin\theta\sqrt{\frac{\mu}{r_{eq}}(2-\frac{q}{r_{eq}})-\frac{\mu}{a}}.
\end{equation}
By expanding the radical term, one finally gets
\begin{equation}
\label{A6}
\frac{dq}{dt}=\xi\sin\theta[B-\frac{\mu}{2Br_{eq}^2}q-\frac{\mu^2}{8B^3r_{eq}^4}q^2],
\end{equation}
where $B$ is already defined in equation \eqref{S3}. The equation (\ref{A6}) is a linear differential equation which can be solved easily. Consequently the equation (\ref{S1}) is obtained as an answer from (\ref{A6}), followed by the relations (\ref{S2})-(\ref{S3}) as definitions in (\ref{S1}).

%%%%%%%%%%%%%%%%%%%%%%%%%%%%%%%%%%%%%%%%%%%%%%%%%%

\end{document}